\documentstyle[preprint,aps,epsf,floats]{revtex}    
\def\bq{\begin{equation}}
\def\eq{\end{equation}}
\def\ba{\begin{eqnarray}}
\def\ea{\end{eqnarray}}

\begin{document}
%
%
\preprint{
\font\fortssbx=cmssbx10 scaled \magstep2
\hbox to \hsize{
\hskip.5in \raise.1in\hbox{\fortssbx University of Wisconsin - Madison}
\hfill$\vtop{   \hbox{\bf MADPH-96-943}
                \hbox{\bf KEK-TH-485}
                \hbox{\bf KEK Preprint 96-24}
                \hbox{May 1996}}$ }
}

\title{\vspace*{.5in}
Probing color-singlet exchange in $Z+2$-jet events at the LHC }
\author{D.~Rainwater$^1$, \ R.~Szalapski$^2$,  \ and D.~Zeppenfeld$^1$\\[3mm]}
\address{
$^1$Department of Physics, University of Wisconsin, Madison, WI 53706\\[2mm]}
\address{$^2$Theory Group, KEK, 1-1 Oho, Tsukuba, Ibaraki 305, Japan}
\maketitle
\begin{abstract}
The purely electroweak process $qq\to qqZ$ (via $t$-channel $\gamma/Z$ 
or $W$ exchange) provides a copious and fairly clean source of color-singlet 
exchange events in $pp$ collisions at the LHC. A judicious choice of 
phase-space region allows the suppression of QCD backgrounds to the 
level of the signal. The color-singlet-exchange signal can be distinguished
from QCD backgrounds by the radiation patterns of additional minijets in 
individual events. A rapidity-gap trigger at the minijet level substantially
enhances the signal versus the background. Analogous features of weak
boson scattering events make $Z+2$-jet events at the LHC an ideal 
laboratory for investigation of the soft-jet activity expected in 
weak-boson scattering events.
\end{abstract}
%
%
\newpage
%
%

\section{Introduction}

The study of weak-boson scattering events and the search for a heavy Higgs 
boson will remain among the most important tasks of the LHC as long as the 
origin of the spontaneous breakdown of the electroweak $SU(2)\times U(1)$ 
gauge symmetry has not been established by experiment. Consequently much work 
has been devoted in recent years on devising methods for the separation of 
weak-boson scattering events, {\em i.e.} the purely electroweak process 
$qq\to qqVV$, from background events like weak-boson pair production
or top-quark decays. One such technique is forward jet tagging, the 
requirement to observe one or both of the two forward quark jets of the 
$qq\to qqVV$ process~\cite{Cahn,BCHP,DGOV}.
However, additional characteristics of the signal must be employed to 
suppress backgrounds.
 
In a weak-boson scattering event no color is exchanged between 
the  initial-state quarks. Color coherence between initial- and final-state
gluon bremsstrahlung then leads to a suppression of hadron production in the 
central region, between the two tagging-jet candidates of the 
signal~\cite{bjgap}. This is in contrast to most background processes
which typically involve color exchange in the $t$-channel and thus lead 
to enhanced particle production in the central region. It was hoped that  
resulting rapidity gaps in signal events (large regions in pseudorapidity 
without observed hadrons) could be used for background suppression. 
Unfortunately, in $pp$ collisions at $\sqrt{s}=14$~TeV at the LHC, the low 
signal cross sections  require running at high luminosity, and then
overlapping events in a single bunch crossing will likely fill a rapidity 
gap even if it is present at the level of a single $pp$ collision. The 
different color structures of signal and background processes can be 
exploited even at high luminosity, however, if one defines 
rapidity gaps in terms of minijets (of transverse momenta in the 20--50~GeV 
range) instead of soft hadrons~\cite{bpz}. 

Sizable background reductions via a minijet veto require the lowering of 
jet-energy thresholds to a range where the probability for additional parton 
emission becomes order unity. In a perturbative calculation the resulting 
condition, $\sigma(n+1\;{\rm jets})\approx \sigma(n\;{\rm jets})$, 
indicates that one is leaving the validity range of fixed-order perturbation
theory, and it becomes difficult to provide reliable theoretical estimates
of minijet emission rates. Gluon emission is governed by very different 
scales in signal as compared to background processes, due to their different 
color structures. Thus a parton shower approach cannot be expected to give 
reliable answers either unless both color coherence and the choice of scale 
are implemented correctly, for which additional information is needed.

\begin{figure}[htb]
\vspace*{-0.1in}
\epsfxsize=5.5in
\epsfysize=3.5in
\begin{center}
\hspace*{0in}
\epsffile{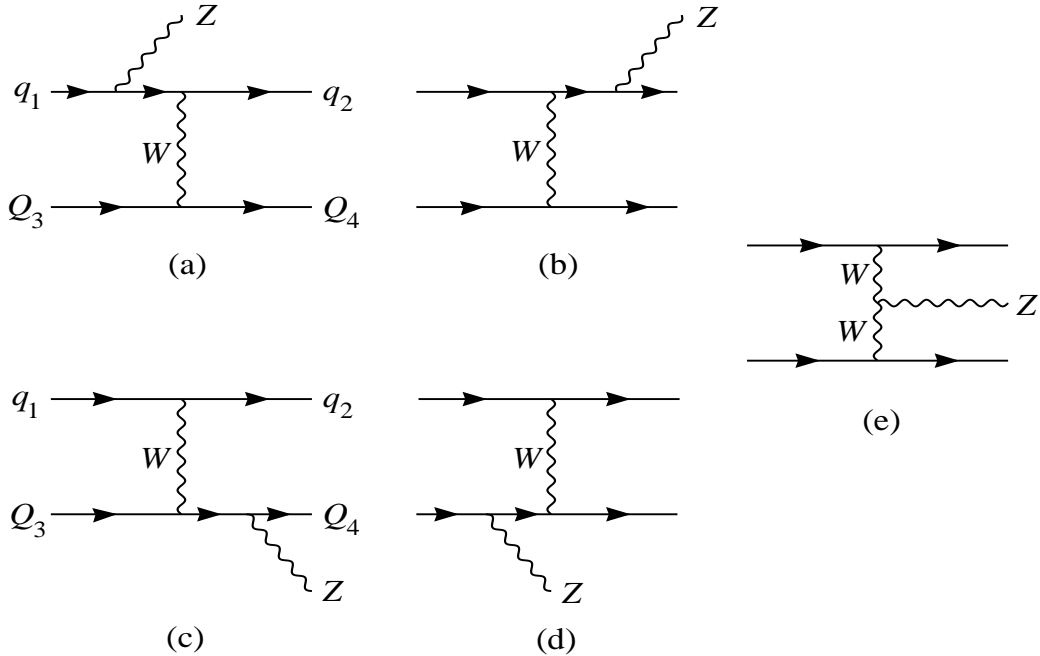}
\vspace*{0.2in}
\caption{
Feynman graphs for $Zjj$ production via charged-current exchange. The 
$WW$-fusion graph (e) simulates weak-boson scattering processes.
\label{fig:zjjfeyn}
}
\vspace*{-0.1in}
\end{center}
\end{figure}

In this paper we describe why and how a different process, $Zjj$ production 
with subsequent $Z\to \ell^+\ell^-$ decay, can be used to answer these 
questions experimentally at the LHC in a region of phase space  very similar 
to the one relevant for weak-boson scattering. The dominant source of $Zjj$ 
events is the ${\cal O} (\alpha_s^2)$ QCD correction to Drell-Yan production.
These events involve color exchange between incident partons, similar to the 
QCD backgrounds to weak-boson scattering events. In addition, there are 
electroweak sources of $Zjj$ events, namely processes of the type $qq\to qqZ$ 
which proceed via color-singlet $\gamma$, $Z$, or $W$ exchange. The 
$W$-exchange process includes the fusion of two virtual $W$'s to a $Z$ boson, 
as shown in Fig.~\ref{fig:zjjfeyn}(e), and thus is very similar to 
Higgs-boson production via weak-boson fusion. 
By tagging the two forward quark jets and requiring a large rapidity 
separation between the two, the 
QCD background can be reduced to the level of the signal, or even below. 
It thus becomes possible to study minijet emission in 
electroweak and QCD $Zjj$ production separately and to obtain the 
necessary experimental information for correct modeling of multiple 
parton emission in $t$-channel color-singlet and color-octet exchange.

Our analysis is based on full tree-level Monte-Carlo programs at 
the parton level. We start out in Section~\ref{sec:two} by describing these
tools. Simulating the minijet emission in $Zjj$ events requires a  
calculation of $Z+3$-jet cross sections. While the QCD 
backgrounds~\cite{HZ,BHOZ,BG} and the $Zjj$ signal process~\cite{CZgap} have 
been available in the literature, we here present a first calculation of  
electroweak $qq\to qqZg$ production (and crossing related processes).
In Section~\protect\ref{sec:three}, using the $Zjj$ programs, we identify 
forward-jet-tagging criteria which lower the QCD backgrounds to approximately 
the level of the signal. We also show how tagging-jet and decay-lepton 
distributions can be used to separate the signal from the background on a 
statistical basis~\cite{CZgap,BZanom}.

Having defined the hard scattering processes to be investigated, we then
turn to the different minijet patterns in signal and background events in 
Section~\ref{sec:four}. Two characteristics differentiate between signal and 
QCD background, the angular distribution of minijets and their typical 
transverse momenta. We discuss the probability for finding minijets in 
hard $Zjj$ events and describe how this probability and the minijet 
multiplicity depend on the phase space region of the hard scattering event. 
Final conclusions are then drawn in Section~\ref{sec:five}.

\section{Calculational Tools }\label{sec:two}

Two aspects of minijet (or soft-gluon) emission in $Zjj$ production need to 
be modelled correctly in order to describe soft-jet activity in these hard
scattering events: the angular distribution of soft emitted partons which
reflects the color coherence specific to the underlying hard scattering event,
and the momentum scale governing soft-gluon emission. Both aspects are 
taken into account correctly by using full tree-level matrix elements for all
subprocesses which contribute to $Zjjj$ production. 

\subsection{The $qq\to qqZ(g)$ signal process}

Our basic signal process is $Z$ bremsstrahlung in quark--(anti)quark scattering
via $W$, $Z$, or photon exchange,
\bq
qQ \to qQZ\;, \qquad  Z\to \ell^+\ell^-\;\;(\ell=e,\;\mu)\; 
\eq
and crossing related subprocesses. Subsequent leptonic $Z$ decay allows 
identification of the signal. The lepton distributions and the tagging of 
the two (anti)quark jets provide a good discrimination 
against QCD backgrounds (see below). In the phase-space region of interest
the  charged-current (CC) 
process of Fig.~\ref{fig:zjjfeyn} dominates over 
neutral-current (NC) exchange, mainly because of the larger coupling of the 
quarks to the $W$ as compared to the photon and $Z$. The $WWZ$ vertex 
in the Feynman graph of Fig.~\ref{fig:zjjfeyn}(e) then leads to a contribution
which very closely resembles  Higgs-boson 
production in weak-boson scattering,
$qq\to qqH$, and thus our signal process becomes a laboratory for studying
QCD aspects of weak-boson scattering.   

We use the results of Ref.~\cite{CZgap} for our calculation of the 
$qq\to qq\ell^+\ell^-$ signal. All CC and NC subprocesses are added, and 
finite $Z$-width effects are included. When requiring a large rapidity
separation between the two quark jets (tagging jets) the resulting large 
dijet invariant mass severely suppresses any $s$-channel processes which
might give rise to the dijet pair. We therefore consider $t$-channel
weak-boson exchange only.
Also note that graphs with $s$-channel electroweak-boson exchange involve 
color exchange between the incident partons and
have a counterpart in the QCD backgrounds to be considered below, but
with electroweak-boson exchange replaced by gluon exchange, 
{\em i.e.} $(\alpha/2{\rm sin}^2\theta_W)^2\approx 2.8\cdot10^{-4}$ replaced 
by $\alpha_s^2\approx 1.4\cdot 10^{-2}$. Thus the electroweak $s$-channel 
processes may be considered as a minor correction to the QCD backgrounds.

In order to determine the minijet activity in signal events we need to evaluate
the ${\cal O}(\alpha_s)$ real parton emission corrections to the signal.
We have performed a first calculation of the ${\cal O}(\alpha^4\alpha_s)$
subprocess 
\vspace*{-0.2in}
\bq
qQ\to qQg\;\ell^+\ell^-
\label{eq:qqZqqg}
\eq
and all crossing related subprocesses. Production of the $\ell^+\ell^-$
pair via $Z$ and $\gamma$ exchange is considered. 
For CC processes, such as $us\to dcg\ell^+\ell^-$, 52 Feynman graphs 
contribute to Eq.~(\ref{eq:qqZqqg}); 
for NC processes 112 Feynman graphs 
need to be included. The resulting amplitudes are evaluated numerically using 
the techniques of Ref.~\cite{HZ,HZor} and have been checked 
against amplitudes generated with MadGraph~\cite{MadGraph}. The cross sections
for the various subprocesses are evaluated and added in a Monte-Carlo 
program\footnote{The code is available upon request from
{\it rain@pheno.physics.wisc.edu}.} 
whose phase-space generator and overall normalization have been tested by
comparing to an analogous $qQ\to qQgH$ generator~\cite{DZ}.

\subsection{The QCD $Zjj(j)$ background}

Given the clean leptonic $Z$ decay signature, the main background to 
electroweak $Z+n$-jet events arises from ${\cal O}(\alpha_s^n)$ real 
emission QCD corrections
to the Drell-Yan process $q\bar q\to Z\to \ell^+\ell^-$.  
For $Zjj$ events these background processes include
\begin{mathletters}
\begin{eqnarray}\label{procback}
q \bar q &\to& g g Z\; , \\
q g &\to& q g Z\; , \\
\noalign{\hbox{\rm or}}
q q &\to& q q Z
\end{eqnarray}
\end{mathletters}
via $t$-channel gluon exchange and all crossing related processes~\cite{Kst}. 
We shall call these processes the 
``QCD $Zjj$'' background. The cross sections 
for the corresponding $Z+3$-jet processes, which we need 
for our modeling of 
minijet activity in the QCD $Zjj$ background, have been calculated in 
Refs.~\cite{HZ,BHOZ,BG}. Similar to the treatment of the signal processes we
use a parton-level Monte-Carlo program based on the work of Ref.~\cite{BHOZ}
to model the QCD $Zjj$ and $Zjjj$ backgrounds. 

For all our numerical results we have chosen 
$\alpha_{QED}=\alpha(M_Z)=1/128.93$,
$M_Z=91.19$~GeV, and $G_F=1.16639\cdot 10^{-5}\;{\rm GeV}^{-2}$, which
translates into $M_W=79.97$~GeV and ${\rm sin}^2\theta_W=0.2310$ when using 
the tree-level relations between these input parameters. The running of the 
strong-coupling constant is evaluated at one-loop order, with 
$\alpha_s(M_Z)=0.12$. MRS~A structure functions~\cite{MRSA} are used 
throughout, and the factorization scale is chosen as the minimal transverse
momentum of a defined jet in the event (see below). For the $qQ \to qQgZ$
signal the scale of the strong coupling constant is taken
to be the minimal transverse momentum of any of the three final 
state partons. For the $Zjj(j)$ QCD backgrounds, with $n=2$ and $n=3$ colored
partons in the final state, the overall strong-coupling constant factors are 
taken as $(\alpha_s)^n = \prod_{i=1}^n \alpha_s(p_{Ti})$, {\em i.e.} the 
transverse momentum of each additional parton is taken as the relevant scale 
for its production, irrespective of the hardness of the underlying scattering
event. This procedure guarantees that the same $\alpha_s^2$ factors are used 
for the hard part of a $Zjj$ event, independent of the number of 
additional minijets, and at the same time the small scales relevant for
soft-gluon emission are implemented.

\section{
$Zjj$ events: electroweak signal and QCD backgrounds}\label{sec:three} 

Before analyzing the minijet activity in signal and background events we 
need to identify the phase-space region for hard scattering events, 
$pp\to ZjjX$ with two hard jets in the final state. In a tree-level
simulation, processes with exactly two final-state partons need to be
considered for this purpose. In the actual experiments this would correspond 
to two-jet inclusive events.  
We are interested in electroweak $Zjj$ production as a model process for
weak-boson scattering. Thus we first need to identify the phase-space region
where the $WW$-fusion graph of Fig.~\ref{fig:zjjfeyn}(e) becomes important. 
This question has been analyzed before for electroweak $Wjj$ production at the
SSC, and we closely follow the procedure outlined in Ref.~\cite{BZanom}. The 
acceptance cuts to be discussed below are chosen with the design of the 
ATLAS and CMS detectors at the LHC in mind~\cite{CMS-ATLAS}.

The leptonic $Z$ decay is a crucial part of the 
signal, and and we therefore 
consider events with two opposite-sign leptons, 
$\ell^+\ell^- = e^+e^-,\;\mu^+\mu^-$, of sufficient transverse momentum,
in the central part of the detector, and well isolated from any jets:
\begin{equation}\label{cut1a}
p_{T\ell}  >  20\; {\rm GeV}\;, \qquad |\eta_\ell|  <  2\; , \qquad
R_{\ell j} = \sqrt{(\eta_\ell-\eta_j)^2 + (\phi_\ell-\phi_j)^2}  >  0.7\;.
\end{equation}
Here $\eta$ denotes the pseudorapidity,  
and $R_{\ell j}$ is the lepton-jet separation in the 
pseudorapidity azimuthal-angle plane. In addition,
the dilepton invariant mass must be consistent with $Z$ decay,
\begin{equation}\label{cut1b}
m_Z-10\;{\rm GeV}\; <\; m_{\ell\ell}\;<\; m_Z+10\,{\rm GeV}\;.
\end{equation}
In the following, 
unless stated otherwise, any parton satisfying the 
transverse momentum, pseudorapidity, and separation requirements
\begin{equation}\label{defjet}
p_{Tj} > 20\, {\rm GeV}\;, \qquad |\eta_j| < 5\;, \qquad R_{jj} >  0.7\;,
\end{equation}
will be called a jet.

\setlength{\unitlength}{0.7mm}
\begin{figure}[t]
\input rotate
\begin{center}
\vspace*{-0.4in}
\hspace*{0in}
\setbox1\vbox{\epsfysize=6.2in\epsffile{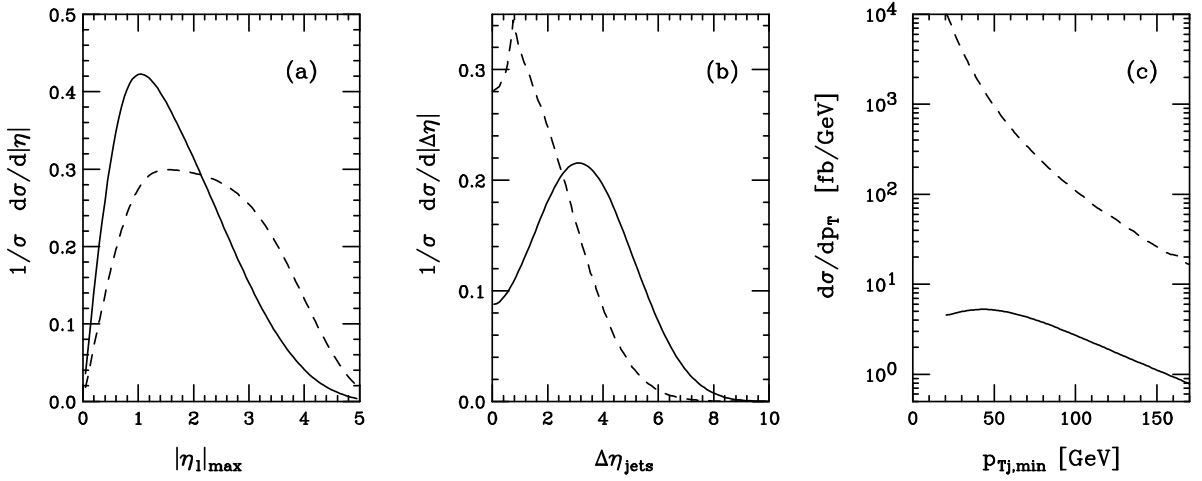}}
\rotr1
\vspace*{0.2in}
\caption{Lepton and jet distributions of signal (solid lines) and background
(dashed lines) $Zjj$ events within the cuts of 
Eqs.~(\protect\ref{cut1a}--\protect\ref{defjet}). Shown are normalized 
distributions of (a) $|\eta_\ell|_{max}$, the maximum 
lepton pseudorapidity, (b) the pseudorapidity separation 
$\Delta\eta_{jets}=|\eta(j_1)-\eta(j_2)|$ of the two jets and (c) the 
differential cross section $d\sigma/dp_{Tj,min}$, where $p_{Tj,min}$ is the 
smaller of the two jet transverse momenta. 
\label{figone}
}
\end{center}
\end{figure}

Event rates after these acceptance cuts are shown in the first row of 
Table~\ref{table1}. Lepton and jet differential 
distributions for the signal (solid lines) and the background (dashed lines)
are shown in Fig.~\ref{figone}. The lepton rapidity distribution of 
Fig.~\ref{figone}(a) shows that signal leptons are more centrally
produced than those in QCD $Zjj$ events. Concentrating on central leptons 
($|\eta_\ell|<2$) does little harm to the signal while reducing the 
background by more than a factor of two. 
A stronger reduction of the background is achieved by exploiting the larger 
pseudorapidity separation of the two jets in the $t$-channel electroweak-boson
exchange of the signal as compared to the QCD background 
(see Fig.~\ref{figone}(b)). Finally, the transverse-momentum distribution of
the softer of the two jets is shown in Fig.~\ref{figone}(c). 

The large jet separation of the signal is typical also for weak-boson 
scattering events, and we therefore require at least three units of 
pseudorapidity between the jet definition cones of the two tagging jets.
In addition, the leptons are required to occupy the pseudorapidity range 
between the two cones, and the two tagging jets must fall into opposite
hemispheres of the detector. With a cone radius of 0.7 for each of 
the jets these conditions can be summarized as
\begin{mathletters}\label{cut2a}
\begin{eqnarray}
|\eta_j^{\rm tag1}-\eta_j^{\rm tag2}|>4.4\;,\qquad && \qquad
\eta_j^{\rm tag1}\cdot\eta_j^{\rm tag2} < 0\;, \\
\eta_j^{\rm tag1}+0.7 < \eta_\ell<\eta_j^{\rm tag2}-0.7 
\qquad &{\rm or}&\qquad
\eta_j^{\rm tag2}+0.7 < \eta_\ell<\eta_j^{\rm tag1}-0.7\;.
\end{eqnarray}
\end{mathletters}
Finally, the jet $p_T$ distributions of Fig.~\ref{figone}(c) suggest
a more stringent transverse-momentum requirement on the tagging jets as 
another means of enhancing the signal with respect to the background. We find 
that a cut at 70~GeV would be optimal for the significance of the signal. 
However, such a high cut would take us well outside the acceptable range
for double jet tagging of weak-boson scattering events. The incident 
longitudinally polarized weak bosons in $qq\to qqH$ events lead to 
substantially lower transverse momenta of the tagging jets than the 
transversely polarized incident $W$'s in the $Zjj$ signal 
(median $p_T\approx 30$~GeV {\em vs.} $\approx 70$~GeV for the softer
of the two tagging jets). Since we want to explore events which are as
similar as possible to longitudinal weak-boson scattering events, we
compromise at
\begin{equation}\label{cut2b}
p_{Tj}^{\rm tag} > 40\, {\rm GeV}\;.
\end{equation}

\begin{table}
\caption{Signal and background cross sections $B\sigma$ for $Zjj(j)$ events 
in $pp$ collisions at $\protect\sqrt{s}=14$~TeV. The two decay 
modes $Z\to e^+e^-,\; \mu^+\mu^-$ are considered. Results are given in units 
of fb after increasingly stringent cuts. The last column gives the ratio of 
signal to background cross section. \label{table1} }
\begin{tabular}{lccc}
\phantom{generic} & $Zjj$ signal &  QCD $Zjj$ background & S/B \\
\hline
generic cuts [Eq.~(\ref{cut1a}--\ref{defjet})]& 
                                              516 & 1.29$\cdot 10^5$& 1:250\\
$+$~forward jet tagging, [Eq.~(\ref{cut2a},\ref{cut2b})]& 
                                             86.6 & 627 & 1:7.2\\
$+$~$m_{jj}>1500$~GeV                 & 44.2 & 87.9 & 1:2.0 \\ 
$+$~$m_{jj}>2500$~GeV, $\Delta\eta_{\ell j}>1.6$  &  10.7  &  6.8  &  1.6:1 \\
$+$~$p_{Tj}>100$~GeV                             &   4.6  &  1.6  &  2.9:1 \\
\end{tabular}
\end{table}

\setlength{\unitlength}{0.7mm}
\begin{figure}[thb]
\input rotate
\begin{center}
\vspace*{-0.4in}
\hspace*{0in}
\setbox1\vbox{\epsfysize=5.8in\epsffile{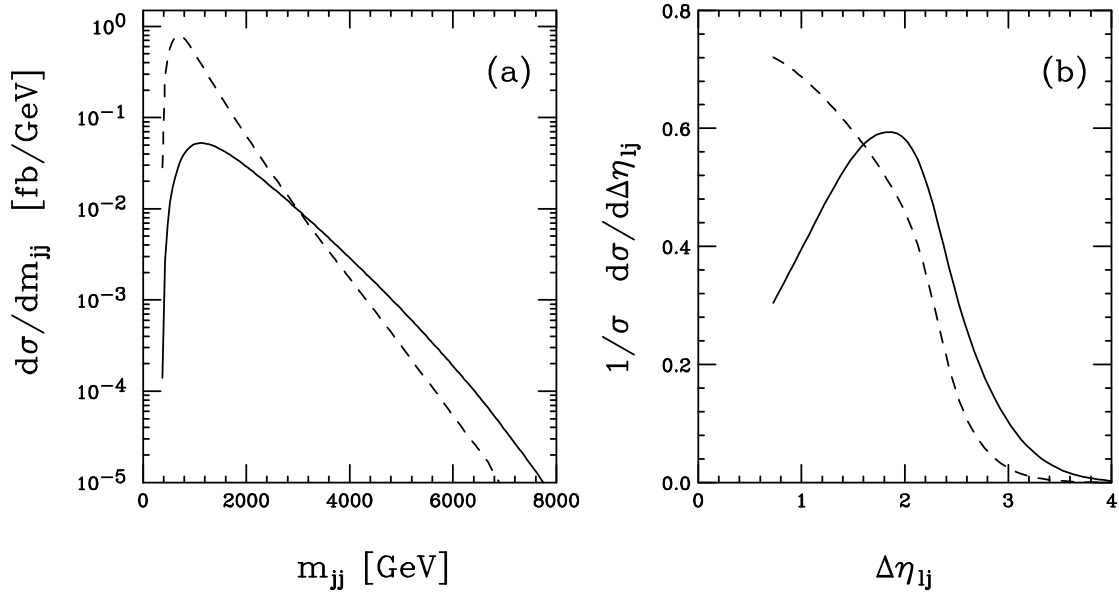}}
\rotr1
\vspace*{0.2in}
\caption{Lepton and jet distributions of signal (solid lines) and background
(dashed lines) $Zjj$ events within the cuts of 
Eqs.~(\protect\ref{cut1a}--\protect\ref{cut2b}). 
Shown are (a) the dijet mass distribution of the two tagging jets,
and (b) the minimal pseudorapidity separation $\Delta\eta_{\ell j}$ 
between any of the leptons and tagging jets. Note that the distribution in 
pseudorapidity separation has been normalized to unit area.
\label{figtwo}
}
\end{center}
\end{figure}
The resulting cross sections, after the cuts of 
Eqs.~(\ref{cut1a}--\ref{cut2b}), are given in the second row of 
Table~\ref{table1}. Distributions in dijet invariant mass and lepton--jet 
separation are shown in Fig.~\ref{figtwo}. These distributions 
clearly show that the QCD $Zjj$ background can be further suppressed with
respect to the signal, {\em e.g.} by increasing $m_{jj}$, the dijet invariant
mass of the two tagging jets, or by requiring a larger minimal separation,
\bq
\Delta\eta_{\ell j} = \min_{\ell,j}\;\{|\eta_\ell -\eta_j^{tag}| \} \; ,
\eq
between the $Z$ decay leptons and the two tagging jets. Cross sections and 
signal to background ratios for three examples of more stringent cuts are 
shown in the last three rows of Table~\ref{table1}; 
it will be possible to prepare event samples
with very different fractions of electroweak- and QCD-induced $Zjj$ events.
The availability of both signal- and background-dominated event samples will 
then allow the study of radiation patterns of minijets in both $t$-channel 
color-singlet exchange events (signal) and in events which are due to color 
exchange between the incident partons (QCD background).

With regard to the separation of signal and QCD background, it should also 
be noted that the calculation of full NLO QCD corrections 
is possible for the $Zjj$ signal with presently available techniques; at
most box diagrams need to be considered at the
one-loop level. Thus it can reasonably be expected that the signal will be 
predictable with good accuracy by the time the LHC can perform these 
measurements. Given the measured event rate and the predicted signal rate 
the composition of the $Zjj$ events should  be known at the 10\% level 
or better. Shape differences of distributions, as in Fig.~\ref{figtwo}, can
then be used to verify the relative composition of event samples.

\section{Radiation patterns of minijets}\label{sec:four}

Having isolated a phase-space region similar to the one populated by 
weak-boson scattering events,
one can use two-jet inclusive $Z$ production events to study the soft-jet
activity in events with or without color exchange in the $t$-channel.
As discussed in Section~\ref{sec:two} we simulate the minijet activity in 
hard $Zjj$ events by generating $Z+3$-parton signal and background events.
In the presence of three jets the tagging jets are now defined as the 
two most energetic jets with $p_T^{tag}>40$~GeV in opposite hemispheres of 
the detector. In the following we are interested in the properties 
of the third or soft parton, which may or may not qualify as a minijet.

\setlength{\unitlength}{0.7mm}
\begin{figure}[t]
\input rotate
\begin{center}
\vspace*{-0.2in}
\hspace*{0in}
\setbox1\vbox{\epsfysize=5.8in\epsffile{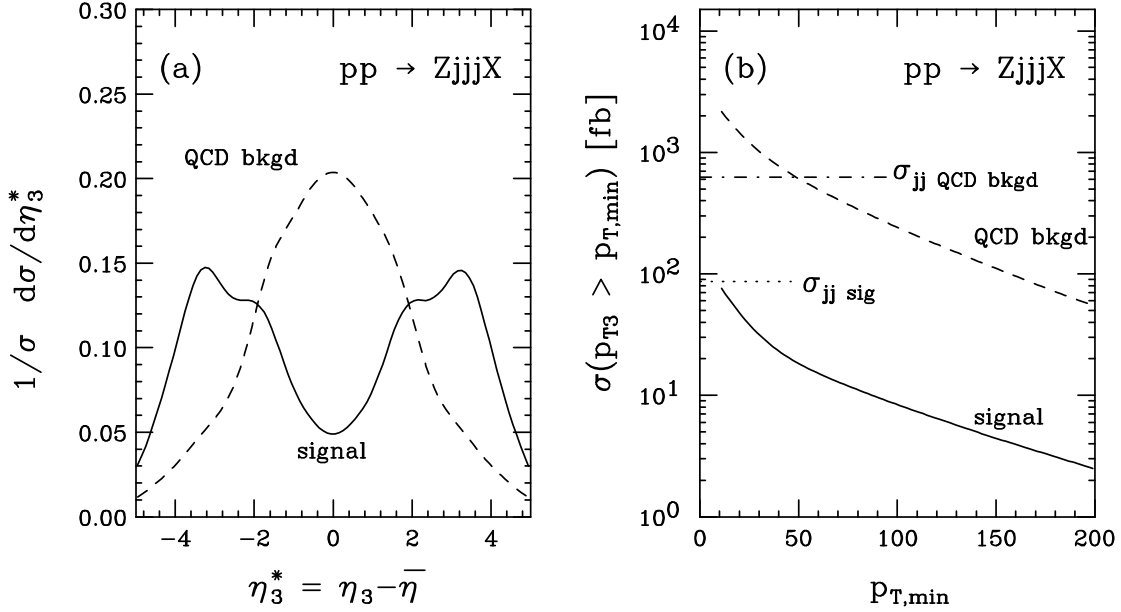}}
\rotr1
\vspace*{0.3in}
\caption{Characteristics of the third (soft) jet in $Zjjj$ signal (solid 
lines) and background (dashed lines) events at the LHC. (a) The 
pseudorapidity $\eta_3^*$ is measured with respect to the center of the two 
tagging jets, $\bar\eta = (\eta_j^{\rm tag1}+\eta_j^{\rm tag2})/2$, and the
distributions are normalized to unit area. (b) Integrated transverse-momentum 
distribution of the third jet, $\sigma(p_{T3}>p_{T,min})$. The acceptance 
requirements of Eqs.~(\protect\ref{cut1a}--\protect\ref{cut2b}) are imposed 
on the two tagging jets. The corresponding cross sections at lowest order,
with two partons in the final state, are indicated for the signal (dotted 
line) and for the background (dash-dotted line).
\label{figthree}
}
\end{center}
\end{figure}

The pseudorapidity and transverse-momentum distributions of this third jet
are shown in Fig.~\ref{figthree}, where the $p_{Tj}$ threshold has been 
lowered to 10~GeV. As expected for $t$-channel color-singlet 
exchange, additional jet activity in the signal is concentrated in the 
forward and backward regions. Color exchange between the incident
partons, as in the case of the QCD background, leads to minijet activity in 
the central region. These differences become particularly pronounced when
measuring the soft jet's rapidity with respect to the center of the two
tagging jets, {\em i.e.} by using the shifted pseudorapidity
\bq
\eta_3^* = \eta_3 - \bar\eta = 
\eta_3 - {\eta_j^{\rm tag1}+\eta_j^{\rm tag2}\over 2}\; .
\eq
The dip in Fig.~\ref{figthree}(a) at $\eta_3^*=0$ is the hallmark of color
coherence in color-singlet exchange~\cite{bjgap,colcoh,fletcher}. 
Beyond this
different angular distribution of the soft-jet activity another striking
difference arises in the transverse-momentum distribution of the third jet; 
the additional jet is substantially harder in the QCD background than in 
the signal. This difference is hardly noticeable in the shape of the 
$p_{T3}$ distribution. It becomes apparent, however, by integrating the $Zjjj$ 
cross section above a given minimum transverse momentum, $p_{T,min}$, of the
non-tagging jet, 
\bq
\sigma_3=\sigma(p_{T3}>p_{T,min}) = \int_{p_{T,min}}^{\infty} 
{d\sigma \over dp_{T3}}\;dp_{T3}\;.
\eq
This integrated three-jet cross section, $\sigma_3$, is shown as a 
function of $p_{T,min}$ and compared to the two-jet cross 
section, $\sigma_2=\sigma_{jj}$, in Fig.~\ref{figthree}(b). 

The number of events with two leptons and two tagging jets, which satisfy
the cuts of Eqs.~(\ref{cut1a}--\ref{cut2b}), will clearly be independent 
of the transverse momentum threshold $p_{T,min}$. At tree level we must
therefore interpret the $Zjj$ cross section $\sigma_2$ as the
two-jet inclusive cross section\footnote{Here and in the following we use 
the term ``$n$-jet inclusive cross section'' to count the number of events 
with $n$ or more jets, {\em i.e.} each event is counted once, independent 
of the jet multiplicity in the event.}. The alternative interpretation 
of $\sigma_2+\sigma_3$ as the two-jet inclusive cross section is unphysical
since $\sigma_3$ can be made arbitrarily large by lowering $p_{T,min}$.

As long as $\sigma_3(p_{T,min})<<\sigma_2$, fixed-order perturbation theory
should be reliable, and we can expect cross sections for four or more jets
to be small. Fig.~\ref{figthree}(b) demonstrates that, for the electroweak 
signal, this perturbative regime covers all $p_T$ thresholds of practical 
interest; $\sigma_3$ saturates the two-jet inclusive cross 
section, $\sigma_2$, at $p_{T,min}({\rm signal})=7.6$~GeV, and this value 
is well below the range where minijets from overlapping events become 
important; at design 
luminosity of ${\cal L}=10^{34}{\rm cm}^{-2}{\rm sec}^{-1}$ a random jet 
of $p_{Tj}\agt 20$~GeV is expected in about 20\% 
of all bunch crossings~\cite{ciapetta}.

The situation is very different for the QCD background. 
Here $\sigma_3\approx \sigma_2$ is reached 
at $p_{T,min}({\rm background})=41$~GeV. Clearly, fixed-order perturbation 
theory is breaking down for $p_{T,min}\alt 70$~GeV and large values 
of $\sigma_4$, $\sigma_5$ {\em etc.} must be 
expected. In the actual experiment multiple minijet emission will appear
in this transverse-momentum range. Thus the $t$-channel color-singlet
exchange of the signal and the color exchange of the QCD background lead to 
dramatically different minijet activity in individual events; 
color-singlet-exchange events will sport a low occupancy of fairly soft jets
in the forward and backward region 
with very little activity in the 
central region, while a typical QCD background event will have several 
minijets of transverse momentum above 20~GeV, predominantly in the central
region, between the the two tagging jets. 

In the analogous case of weak-boson scattering events the same pattern 
arises, and a veto 
on central minijets can be used to suppress the 
backgrounds~\cite{bpz}. The efficiency of a minijet veto can be tested 
experimentally at the LHC using 
the $Zjj$ events discussed here. The precise
definition of a minijet veto will depend on detector performance, multiplicity
of minijets from overlapping events~\cite{ciapetta}, and detailed signal and 
background characteristics. Given the characteristics of signal and background
$Zjjj$ events discussed above, the veto region may be defined as the 
pseudorapidity range between the tangents to the two tagging jets, and as jet 
transverse momenta above a minimal value, $p_{T,\rm veto}$,
\begin{mathletters}\label{eq:veto}
\begin{eqnarray}
p_{Tj}^{\rm veto} & > & p_{T,\rm veto}\;, \label{eq:ptveto} \\
{\rm min}\;\{\eta_j^{tag1},\eta_j^{tag2}\}+0.7 & < & \eta_j^{\rm veto} 
< {\rm max}\;\{\eta_j^{tag1},\eta_j^{tag2}\}-0.7\; , \label{eq:etaveto}
\end{eqnarray}
\end{mathletters}
and we will use this definition as an example in the following.

Jets with transverse momentum in the 20~GeV range should be observable in 
hard events at the LHC~\cite{ciapetta}, 
and perhaps even lower thresholds are 
possible at luminosities below ${\cal L}=10^{33}{\rm cm}^{-2}{\rm sec}^{-1}$.
Since, for the QCD background, this $p_T$ range is below the validity region 
of fixed-order QCD, we need to resort to some modeling in order to estimate
the probability for multiple minijet emission. Any model should preserve 
the two salient features of the QCD 
matrix-element calculation: color coherence
as reflected by the angular distributions of Fig.~\ref{figthree}(a) and the 
different $p_T$ scales for extra parton emission that we have found for the 
signal and the background. 

\begin{figure}[htb]
\input rotate
\begin{center}
\vspace*{-0.2in}
\hspace*{0in}
\setbox1\vbox{\epsfysize=4in\epsffile{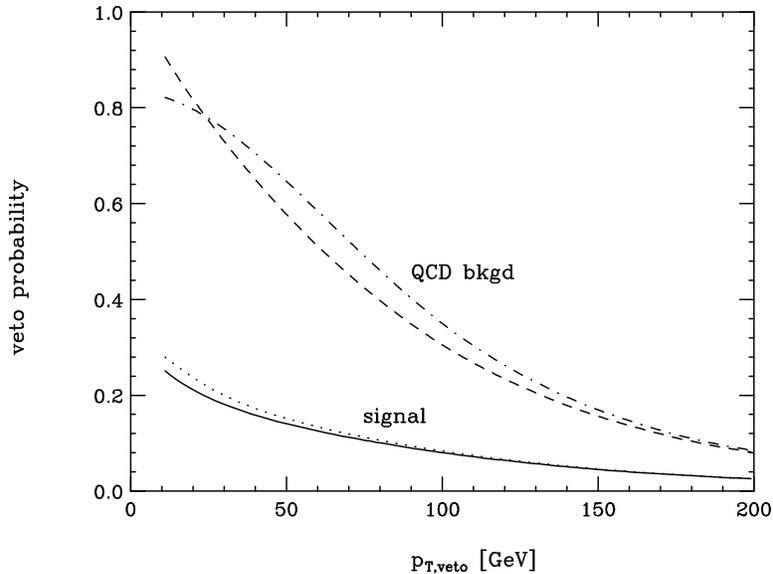}}
\rotr1
\vspace*{0.2in}
\caption{Probability to find a veto jet with transverse momentum above
$p_{T,\rm veto}$ and in the pseudorapidity range of 
Eq.~(\protect\ref{eq:etaveto}) in signal and background events within the 
cuts of Eqs.~(\protect\ref{cut1a}-\protect\ref{cut2b}). The solid
(signal) and dashed (background) curves are 
obtained with the exponentiation ansatz of Eq.~(\protect\ref{eq:expon})
while the truncated shower approximation yields the dotted curve for the 
signal and dash-dotted curve for the QCD background.
\label{figfour}
}
\end{center}
\end{figure}

In the following we use the two models discussed in Ref.~\cite{bpz}. The 
first one is provided by the ``truncated shower approximation'' 
(TSA)~\cite{pps}. When several soft gluons are emitted in a hard 
scattering event their transverse momenta tend to cancel, leading to a 
regularization of the small $p_T$ singularity which is present when 
considering single-parton emission only. In the TSA these effects are 
simulated by replacing the tree-level three-jet 
differential cross section, $d\sigma_3^{\rm TL}$, with
\begin{equation}\label{eq:tsa}
d\sigma_3^{\rm TSA}=d\sigma_3^{\rm TL}
\left(1-e^{-p_{T3}^2/p_{TSA}^2}\right)\;.
\end{equation}
Here the parameter $p_{TSA}$ is chosen to correctly reproduce the 
tree-level two-jet cross section, $\sigma_2$, within the cuts of 
Eqs.~(\ref{cut1a}-\ref{cut2b}), {\it i.e.} $p_{TSA}$ is fixed by the 
matching condition
\bq
\sigma_2 = \int_0^\infty {d\sigma_3^{\rm TSA}\over dp_{T3}} dp_{T3}\; .
\eq
This is achieved by setting $p_{TSA}=10.5$~GeV
for the $Zjj$ signal and $p_{TSA}=72$~GeV for the QCD $Zjj$ background. 
The much larger value for the latter again reflects the higher intrinsic 
momentum scale governing soft-gluon emission in the QCD background. This 
difference would be enhanced even more by requiring larger dijet invariant 
masses for the two tagging jets, as in the final two rows of 
Table~\ref{table1} (see also below). Using $d\sigma_3^{\rm TSA}$ as a model
for additional jet activity we find the probabilities of Fig.~\ref{figfour}
(dotted and dash-dotted curves) for emission of a third, soft parton into
the veto region of Eq.~(\ref{eq:veto}).

\setlength{\unitlength}{0.7mm}

In the TSA only one soft parton is generated, with 
a finite probability to be produced outside the veto region of 
Eq.~(\ref{eq:etaveto}). The veto probability will therefore never 
reach 1, no matter how low a $p_{T,\rm veto}$ is allowed. At small 
values of $p_{T,\rm veto}$ we underestimate the veto probability because 
the TSA does not take into account multiple parton emission. In the soft 
region gluon emission dominates, and one may assume that this soft-gluon 
radiation approximately exponentiates, {\em i.e.} the probability $P_n$ for 
observing $n$ soft jets in the veto region is given by a Poisson distribution,
\bq\label{eq:expon}
P_n = {\bar n^n\over n!}\; e^{-\bar n} \;,
\eq
with
\bq
\bar n = \bar n(p_{T,\rm veto}) = {1 \over \sigma_2}\; 
\int_{p_{T,\rm veto}}^{\infty} dp_{T3}\; 
{d\sigma_3 \over dp_{T3}}\; ,
\eq
where the unregularized three-parton cross 
section is integrated over the veto 
region of Eq.~(\ref{eq:veto}) and then normalized to the $Zjj$ cross 
section, $\sigma_2$. We will call this model the ``exponentiation model''.
A rough estimate of multiple emission effects is thus provided 
by using
\begin{equation}\label{Pvetoexp}
P_{exp}(p_{T,\rm veto}) = 1-P_0 = 1- e^{-\bar n(p_{T,\rm veto})} 
\end{equation}
for the veto probability. The resulting curves are the solid and dashed lines
in Fig.~\ref{figfour}. In spite of the approximations made, both models 
agree qualitatively on the much larger probability to observe additional
minijets in the QCD background as compared to the $Zjj$ signal. 

\begin{figure}[t]
\input rotate
\begin{center}
\vspace*{-0.2in}
\hspace*{0in}
\setbox1\vbox{\epsfysize=5in\epsffile{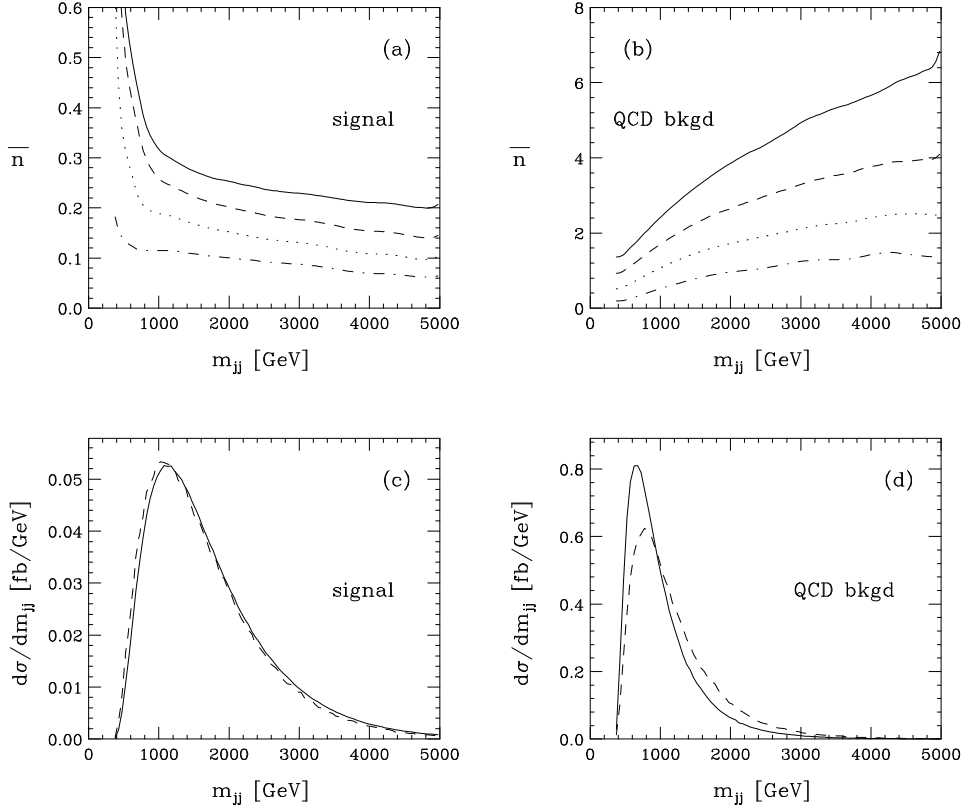}}
\rotr1
\vspace*{0.2in}
\caption{
Average minijet multiplicity $\bar n = \sigma_3(p_{T,min})/\sigma_2$ in 
the veto region of Eq.~(\protect\ref{eq:veto}) as a function of the invariant 
mass of the two tagging jets, $m_{jj}$, for four different transverse 
momentum thresholds of the third jet: $p_{T,\rm veto}=10$~GeV (solid line), 
20~GeV (dashed line), 40~GeV (dotted line), and 80~GeV (dash-dotted line). 
Results are shown for (a) the $Zjj$ signal and (b) the QCD background. Below 
each, $d\sigma/dm_{jj}$ is shown as determined with $Zjj$ tree-level matrix 
elements (solid lines) and by using the truncated shower approximation 
(dashed lines).
\label{fig:nbar.mjj}
}
\vspace*{-0.05in}
\end{center}
\end{figure}

Within the exponentiation model, $\bar n = \sigma_3/\sigma_2$ represents the 
average multiplicity of minijets in the central region, between the two 
tagging jets. Even if the exponentiation model is of limited accuracy only,
the ratio of three- to two-jet tree-level cross sections gives the 
best perturbative estimate available of the minijet activity in $Zjj$ events.
One finds that the average minijet multiplicity depends strongly on the 
hardness of the underlying $Zjj$ event. 
In Fig.~\ref{fig:nbar.mjj}(a,b) the dependence of $\bar n$ on the dijet 
invariant mass of the two tagging jets is shown for both the signal and 
the QCD background, for four values of the transverse-momentum cut on 
minijets: $p_{T,\rm veto}=10$~GeV (solid line), 
20~GeV (dashed line), 40~GeV (dotted line), and 80~GeV (dash-dotted line). 
The differential cross sections $d\sigma/dm_{jj}$ are shown in 
Fig.~\ref{fig:nbar.mjj}(c,d), allowing an assessment of the relative 
importance of regions of different $\bar n$.

Except for the threshold region, where kinematical effects of
additional minijet emission are most important, the minijet multiplicity
in signal events is below 25\% everywhere, and thus 
fixed-order perturbation 
theory should be reliable. Perhaps somewhat surprisingly, the minijet activity
of the signal decreases with increasing $m_{jj}$. This effect can be traced to
the relative contribution of gluon-initiated $gQ \to q\bar q QZ$ events
as compared to events with soft gluons in the final state, $qQ\to qQ Zg$.
The splitting process $g\to q\bar q$ 
has a much higher probability to produce a semi-hard 
central jet than gluon radiation in $t$-channel color-singlet exchange. In the
latter case color coherence between initial- and final-state radiation forces 
the gluon jet into the forward and backward regions~\cite{bjgap}, 
and in addition the transverse-momentum spectrum of the produced gluon is 
much softer than that of the additional quark jet in $g\to q\bar q$ splitting. 
By themselves, $qQ \to qQZg$ events would 
produce an essentially flat minijet multiplicity distribution, which, 
for $p_{T3}>20$~GeV, varies between $\bar n= 0.16$ and $\bar n=0.12$ over 
the entire $m_{jj}$ range shown in Fig.\ref{fig:nbar.mjj}(a). Since 
high $m_{jj}$ events populate the large Feynman-$x$ region, where the 
valence-quark distributions dominate, the pattern expected for final-state 
gluons is found when concentrating on the high-invariant-mass region.

The situation is entirely different for high-mass QCD background events, which
are dominated by $t$-channel gluon exchange. The relevant scale for the 
acceleration of color charges and, hence, the emission of soft gluons, is 
set by $m_{jj}$, and, as a result, the minijet multiplicity in 
Fig.~\ref{fig:nbar.mjj}(b) increases substantially with rising invariant 
mass of the two tagging jets. In addition the expected minijet multiplicity 
is about an order of magnitude higher in QCD background events than in 
the $Zjj$ signal. This different dependence of $\bar n$ on $m_{jj}$ has an 
intriguing consequence: plotting the minijet multiplicity distribution 
of $Zjj$ data in increasingly higher $m_{jj}$ bins, one expects a clear 
separation to develop between the signal events, which will be concentrated 
at zero minijet multiplicity, and the QCD background events, which will 
populate the high-multiplicity region.

\setlength{\unitlength}{0.7mm}
\begin{figure}[t]
\input rotate
\begin{center}
\hspace*{0in}
\setbox1\vbox{\epsfysize=5in\epsffile{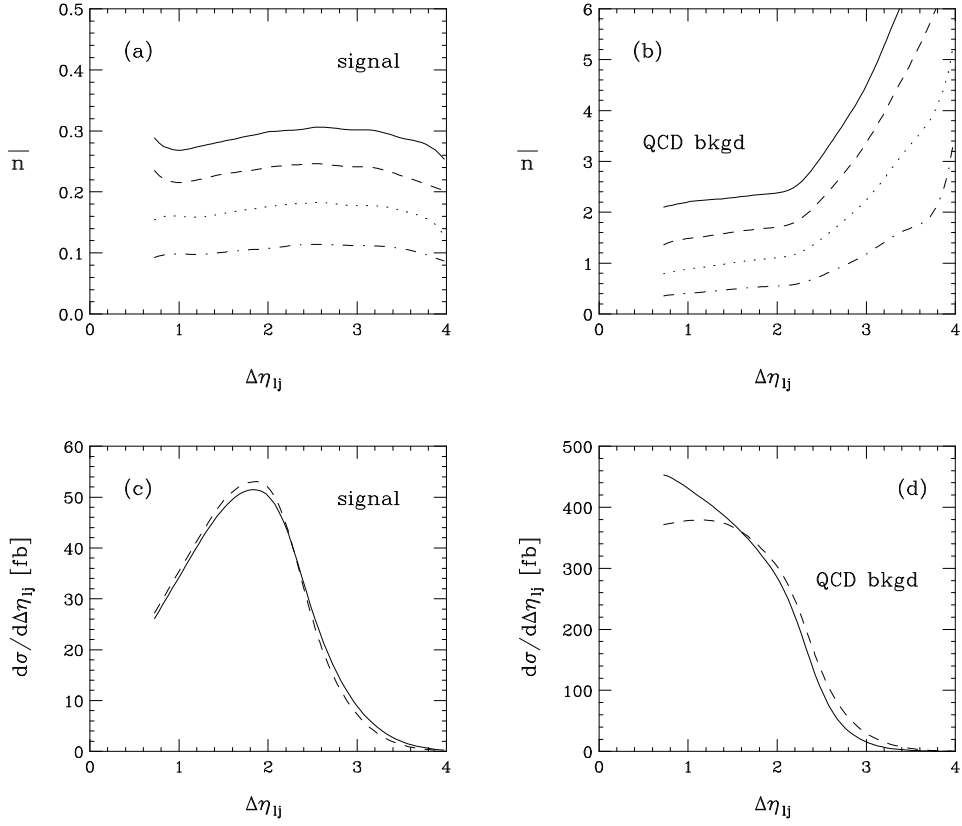}}
\rotr1
\vspace*{0.4in}
\caption{
Same as Fig.~\protect\ref{fig:nbar.mjj} but for the dependence of the 
minijet activity on the minimal separation $\Delta\eta_{\ell j}$ of the 
$Z$-decay leptons from the two tagging jets. See text for details.
\label{fig:nbar.etalj}
}
\end{center}
\end{figure}

The emission of additional partons depends above all on the energy scale of
the underlying hard process. The minijet multiplicity shows much less 
variation with angular variables. One example is given in 
Fig.~\ref{fig:nbar.etalj} where the dependence of $\bar n$ on the minimal 
separation $\Delta\eta_{\ell j}$
%
%
of the $Z$-decay leptons from the two tagging jets is shown. For separations
below $\approx 2.2$, where both signal and background cross sections are 
sizable, $\bar n$ is essentially independent of $\Delta\eta_{\ell j}$. Given
the different shapes of $d\sigma/d \Delta\eta_{\ell j}$, a statistical 
separation of signal and QCD background may be possible for events with any 
given number of minijets. This would allow the independent measurement of the
minijet multiplicity distributions for $t$-channel color-singlet exchange and
QCD background events, thus providing important input
for weak-boson scattering events and their backgrounds in basically the same
kinematical regime.

\section{Discussion}\label{sec:five}

The production of $Zjj$ events at the LHC, with one forward and one backward 
tagging jet which are widely separated in pseudorapidity, provides an ideal
testing ground for the study of $t$-channel color-singlet exchange events.
The electroweak process $qq\to qqZ$ possesses all the relevant characteristics
of weak-boson scattering events. However, it is easily identifiable
via the leptonic $Z$-decay mode, and it enjoys a large production cross 
section (of order 30--80 fb) in a phase-space region where QCD backgrounds 
are of comparable size. Even when operating the LHC at 10\% of the design 
luminosity ({\em i.e.} collecting 10~fb$^{-1}$ per year) a combined signal 
and background sample of more than 1000 events will be available to find 
differences between events with and without color exchange in the $t$-channel.
By varying the machine luminosity, the minijet background from overlapping 
events in the same bunch crossing, and methods for its suppression, can be 
studied at the same time~\cite{ciapetta}.

Color-singlet exchange in the $t$-channel, as encountered in Higgs-boson 
production by weak-boson fusion and in our $Zjj$ signal, leads to 
soft-minijet activity which differs strikingly from that expected for
the QCD backgrounds in at least two respects. First, $t$-channel 
color-singlet exchange leads to soft gluon emission mainly in the forward 
and backward regions, between the beam directions and the forward tagging 
jets~\cite{bjgap,fletcher}. The central region
between the two tagging jets, which also contains the two $Z$-decay leptons, 
remains largely free of minijets. For the backgrounds, 
$t$-channel color exchange leads to minijet emission mainly in the central
region~\cite{bpz}. 

A second distinction is the typical transverse momentum of the produced 
minijets. Extra gluon emission in $Zjj$ production is suppressed by a
factor $f_s=\alpha_s {\rm ln}\; (Q^2/p_{T,{\rm min}}^2)$, where $Q$ is 
the typical scale of the hard process and $p_{T,{\rm min}}$ is the minimal 
transverse momentum required for a parton to qualify as a minijet. 
The jet-transverse-momentum scale below which multiple minijet emission must 
be expected is set by $f_s\approx 1$. The hard scale $Q$ is set by the 
momentum transfer to the color charges in $Zjj$ production. For the signal 
no color is exchanged, and hence the color charges are accelerated by the 
same amount as the incoming (anti)quarks. Hence, $Q$ is related to the 
average $p_T$ of the two tagging jets and is of order 100~GeV only. For the 
background processes, on the other hand, color is exchanged in the 
annihilation of the initial quarks and/or gluons. Therefore the 
momentum transfer to the color charges is of the order of the dijet 
invariant mass of the two tagging jets and  is in the TeV range.
As a result 
multiple minijet emission becomes important in background processes in the 
20--50~GeV $p_T$ range whereas the corresponding scale for the signal is 
of order a few GeV only~\cite{bpz}. 

These qualitative arguments are directly confirmed by our perturbative 
analysis. We find, for example, that minijet emission
in the QCD background increases with the invariant mass of the two 
tagging jets, and that it occurs with much higher probability than for the 
signal even though the transverse momenta of the tagging jets are 
somewhat larger in the electroweak $Zjj$ signal than in the QCD background.
This pattern is naturally explained by taking $Q=m_{jj}$ for the background
and $Q=p_{Tj}^{tag}$ for the signal. 

A precise modeling of multiple minijet emission in hard QCD processes is 
beyond the scope of the present paper. However, any Monte-Carlo program which 
addresses this question should incorporate the above findings and agree with
the fixed-order perturbation-theory results at sufficiently large minijet
transverse momenta. $Zjj$ events at the LHC can then be used to fine-tune
the Monte Carlos in the low $p_T$ range. 

Because of its intrinsically small scale, fixed-order QCD should be reliable 
for the signal process down
to minijet transverse momenta in the 10--20~GeV range, a point which can 
be tested experimentally by comparing the rate of low minijet multiplicity 
$Z+2$-jet inclusive events with the signal 
predictions (to NLO if available by the time the LHC starts running). 

Minijet activity in high-mass QCD events is most easily probed by studying
two-jet inclusive events (without an accompanying $Z$ boson), as is already 
possible now, at the Tevatron~\cite{CDFmj,sz}. However, such events will 
have a composition of quark- and gluon-initiated subprocesses different from
backgrounds to weak-boson scattering events. A test run to prepare for
the latter can be performed with the $Zjj$ QCD backgrounds studied here.
Samples of such events can be prepared either by subtracting the
known electroweak $Zjj$ production cross section, by relying on differing 
shapes of kinematical distribution at the $Z+2$-jet level, or by going to 
relatively low dijet-mass regions where the QCD $Zjj$ background is the 
dominant source of $Zjj$ events. 

Most likely a combination of all of these will be needed to obtain 
an understanding of the minijet activity in hard scattering events 
at a quantitative level. This knowledge can then be used to devise a 
minijet trigger for the Higgs-boson search at the LHC. Our findings 
here indicate that a minijet veto should work not only for the heavy 
Higgs-boson search, where the production of a high-mass system lets one 
expect strong gluon radiation in background events, but also for the 
production of light weak bosons via $WW$ or $ZZ$ fusion. 
This weak boson need not be the $Z$ studied here but could be
an intermediate-mass Higgs boson. The use of a minijet veto appears to be 
a promising technique for the entire Higgs mass range, from the 100~GeV
range of supersymmetric models to the TeV scale.

\acknowledgements
We thank Tim Stelzer for his aid in testing our matrix elements by a 
comparison with MadGraph. This research was supported in part by the 
University of Wisconsin Research Committee with funds granted by the 
Wisconsin Alumni Research Foundation and in part by the U.~S.~Department 
of Energy under Contract No.~DE-FG02-95ER40896.

\newpage             

\end{document}